      [1999/12/01 v1.4c Il Nuovo Cimento]

\documentclass{cimento}

\usepackage{cite}
\usepackage{amsmath}

\newcommand{\NP}[1]{Nucl.\ Phys.\ {\bf #1}}
\newcommand{\PL}[1]{Phys.\ Lett.\ {\bf #1}}

\newcommand{\PR}[1]{Phys.\ Rev.\ {\bf #1}}

\newcommand{\psl}{\mbox{$p$\hspace{-0.4em}\raisebox{0.1ex}{$/$}}}
\newcommand{\ksl}{\mbox{$k$\hspace{-0.4em}\raisebox{0.1ex}{$/$}}}

\usepackage{graphicx}
\title{Infrared singularities in QCD amplitudes}     
\author{E. Gardi\from{Ed} \atque
L.~Magnea\from{To}}
\instlist{\inst{Ed} 
School of Physics, The University of Edinburgh,\\
Edinburgh EH9 3JZ, Scotland, UK
  \inst{To} 
Dipartimento di Fisica Teorica, Universit{\`a} di Torino, and\\
INFN, Sezione di Torino, Via P. Giuria 1, I-10125 Torino, Italy
}
\PACSes{
\PACSit{11.15.-q}{Gauge field theories}
\PACSit{12.38.-t}{Quantum chromodynamics}
\PACSit{12.38.Cy}{Summation of perturbation theory} 
}

\begin{document}

\maketitle

\begin{abstract}
We review recent progress in determining the infrared singularity structure of on-shell scattering amplitudes in massless gauge theories.
We present a simple ansatz where soft singularities of any scattering amplitude of massless partons, to any loop order, are written as a sum over colour dipoles, governed by the cusp anomalous dimension. We explain how this formula was obtained, as the simplest solution to a newly-derived set of equations constraining the singularity structure to all orders. We emphasize the physical ideas underlying this derivation: the factorization of soft and collinear modes, the special properties of soft gluon interactions, and the notion of the cusp anomaly. Finally, we briefly discuss potential multi-loop contributions going beyond the sum-over-dipoles formula, which cannot be excluded at present.
\end{abstract}

\section{The role of infrared singularities}

Understanding soft and collinear singularities is essential for the application 
of QCD to collider physics.  Indeed, cross section calculations beyond tree level 
involve intricate cancellations of such singularities in the sum over final states. 
A detailed understanding of the singularities is therefore a prior condition to 
making precise predictions. Furthermore, knowing the singularities, one can 
resum the dominant radiative corrections to all orders, greatly improving the 
accuracy of the prediction.

Beyond their immediate significance to phenomenology, infrared singularities 
open a window into the all-order structure of perturbation theory. They admit
a simple, iterative structure, which is common to all gauge theories. Understanding 
this structure is an important step towards understanding scattering amplitudes in
gauge theories in general. As an example, recent progress in studying scattering
amplitudes in the maximally supersymmetric (${\cal N} = 4$) Yang-Mills theory 
in the planar limit~\cite{Alday:2007hr,Korchetal,Bern:2008ap,Bern:2005iz,
Alday:2008yw}, has demonstrated that in that case the iterative structure of 
the amplitude persists in its finite parts. 
Moreover, for the first time a bridge was formed between the weak coupling 
expansion and the strong coupling limit. In these studies the infrared singularity
structure had a key role. In particular, the cusp anomalous dimension 
$\gamma_K (\alpha_s)$~\cite{Polyakov:1980ca,Korchemsky:1985xj,Ivanov:1985np,Korchemsky:1987wg,Korchemsky:1988hd,Korchemsky:1988si}, which, 
as we shall see, governs soft singularities in any scattering amplitude, was shown
to have an important role also in determining the finite parts of the amplitude. 
Today $\gamma_K (\alpha_s)$ is the best understood anomalous dimension, 
at both weak~\cite{Moch:2004pa} and strong 
coupling~\cite{Alday:2007hr,Beisert:2005fw,Beisert:2006ez,Basso:2007wd}.
As shown by this example, there is a strong theoretical incentive to gain full understanding of the singularity structure of scattering amplitudes. Let us 
now discuss the more pragmatic motivation aiming at precision collider
phenomenology. 

\subsection*{Cross--section calculations beyond tree level}

The very fact that gauge-theory amplitudes are plagued by long-distance
singularities while the corresponding cross sections are finite, makes the 
determination of these singularities essential. The cancellation of infrared 
singularities in cross sections takes place upon summing over degenerate 
states, as originally shown in QED by Bloch and Nordsieck~\cite{Bloch:1937pw},
and later proven in~\cite{Kinoshita:1962ur,Lee:1964is}. Virtual gluons generate
singularities in amplitudes owing to the integrations over loop momenta, which 
extend over regions where the gluons are soft or collinear with any of the hard 
partons --- this puts some internal propagators on shell, leading to singularities. 
In contrast, real emission diagrams are finite, but singularities appear upon performing phase-space integrations over regions where the emitted partons 
become soft or collinear with other partons.    
The physical cross section is a sum of these two contributions, which can 
be separately computed in dimensional regularization. Schematically, using 
$D = 4 - 2\epsilon$ dimensions with $\epsilon<0$, one finds cancellations 
of the form
\begin{equation}
\label{real_virt_canc}
 \qquad \underbrace{{\frac{1}{\epsilon}_{\,\,}}_{\,}}_{\rm virtual}\,\,\,\,
 + \, \underbrace{
 (Q^2)^{\epsilon} \int_0^{m_{\rm jet}^2} \frac{d k^2}{(k^2)^{1 + \epsilon}}
 }_{\rm real} \,\, \qquad \quad \Longrightarrow \,\,\,\,\,\, 
 \ln (m_{\rm jet}^2/Q^2) \, ,
\end{equation}
where $Q^2$ represents a hard energy scale, {\it e.g.} the squared 
centre-of-mass energy $s$, while $m_{\rm jet}^2$ represents the 
phase-space limit in the integration of the radiated gluon, which depends 
on the observable considered, {\it e.g.} a jet mass. 
Because of their different origin, these singularities render any calculation of scattering cross sections beyond tree level highly non-trivial. 

At the one-loop order we have a complete understanding of these singularities. This forms the basis for general subtraction algorithms, for example based on a colour dipole picture~\cite{Catani:1996jh}, rendering the phase--space integration finite. 
The possibility to perform such local subtraction has been invaluable to practical 
cross--section calculations. Present day collider phenomenology requires 
computations of multi-leg processes in general kinematics, in order to allow for 
maximal flexibility in the application of kinematic cuts dictated by the search 
strategies and experimental needs. This leads to complicated phase-space
integrations, which can only be done numerically. Thus, a local subtraction of 
the singularities --- which guarantees finite integrals --- is an absolute necessity.
General subtraction algorithms do not exist yet at the multi-loop level and their 
development is of prime importance to precision computations. The first step 
in this direction is the determination of the singularity structure of amplitudes, 
the subject of the present talk.

\subsection*{Resummation}

Beyond fixed-order cross-section calculations, infrared singularities also provide 
the key to resummation of soft and collinear gluon radiation. Singular contributions
cancel between real and virtual corrections, but, as shown schematically in 
eq.~(\ref{real_virt_canc}), a residual logarithm survives. 
These logarithmically--enhanced corrections (Sudakov logarithms) are the 
dominant radiative corrections for many cross sections. In particular, these
corrections are parametrically leading when the relevant scales are
far apart --- for example in eq.~(\ref{real_virt_canc}) when $m_{\rm jet}^2
\ll Q^2$. Whenever the logarithm becomes as large as the inverse power of 
the coupling, it spoils the converges of the expansion, and corrections involving 
powers of $\alpha_s \, \ln( m_{\rm jet}^2/Q^2)$ need to be resummed to 
all orders. The situation is complicated by the fact that, due to overlapping 
soft and collinear divergences, each order in perturbation theory can give rise 
to two logarithms, yielding $\alpha_s\, \ln^2( m_{\rm jet}^2/Q^2)$. In this 
case, resummation is necessary already for $\ln( m_{\rm jet}^2/Q^2)\sim 
1/\sqrt{\alpha_s}$. Because these logarithms are all generated by the 
singularities in the amplitude, which always exponentiate, higher powers of 
the logarithms at any order in the coupling can be predicted based on the 
singular terms in the first few orders in the loop expansion. This is a key 
ingredient for resummation.

The most widely used applications of this picture are parton--shower event 
generators, which implement Sudakov resummation with leading logarithmic 
accuracy, keeping complete kinematic information on the generated final state. 
To achieve better precision one typically resorts to a more inclusive approach.
Indeed, it has been repeatedly demonstrated in a variety of applications, 
e.g.~\cite{Gardi:2001ny,Cacciari:2002xb,Bozzi:2007pn,Ahrens:2008nc,
Andersen:2006hr,Andersen:2005mj}, that precise predictions can be obtained 
in kinematic regions that are characterized by a large hierarchy of scales upon 
performing Sudakov resummation, provided one gains sufficient control of
subleading logarithms and related power corrections.

The theory of Sudakov resummation is especially well developed in inclusive
observables, where the hard scattering process involves just two coloured 
partons\cite{Sterman:1986aj,Catani:1989ne,Contopanagos:1996nh,Laenen:2000ht,
Bozzi:2007pn,Ahrens:2008nc,Gardi:2007ma,Gardi:2006jc}. Such processes are 
characterized by a single or a double hierarchy of scales. Examples of the first 
category include deep-inelastic structure functions at large Bjorken 
$x$~\cite{Gardi:2002xm}, and Drell-Yan or Higgs production near partonic threshold, or at small transverse momentum~\cite{Bozzi:2007pn,Ahrens:2008nc}.
Examples of the second include event--shape 
distributions~\cite{Catani:1992ua,Gardi:2001ny}, heavy quark production~\cite{Kidonakis:1999ze,
Cacciari:2002xb}, and inclusive meson decay spectra~\cite{Andersen:2006hr,
Andersen:2005mj}. The Sudakov factor in processes involving two (incoming 
or outgoing) partons, may be written in the generic form
\begin{align}
\label{Sud}
{\rm Sud}(m^2,N)=\exp\left\{C_i\int_0^1\frac{dr}{r} \Big[\underbrace{(1-r)^{N-1}}_{\rm real}\,\,\underbrace{-1}_{\rm virtual}\Big] R(m^2,r)\right\},
\end{align}
where $C_i=C_F$ or $C_A$ depending on the colour representation of the hard 
partons (fundamental or adjoint), and the radiator is given by
\begin{align}
\label{Rad}
C_i\frac{R(m^2,r)}{r}=-\frac{1}{r}\,\left[\int_{{ r^2m^2}}^{{ r m^2}}\frac{dk^2}{k^2}{\gamma_K}\left(\alpha_s(k^2)\right)+ { 2 {\cal B}}\left(\alpha_s({ r m^2})\right)-
{ 2{\cal D}}\left(\alpha_s({ r^2 m^2})\right)\right]\,.
\end{align}   
These two equations summarize, in a compact way, the form of 
logarithmically-enhanced terms in a typical infrared-safe cross section, to all 
orders in perturbation theory.
This simple structure is a consequence of factorization, namely the fact 
that soft and jet subprocesses decouple from the hard interaction and are 
mutually incoherent.        
Eq.~(\ref{Sud}) incorporates the cancellation between real and virtual singularities 
anticipated in (\ref{real_virt_canc}). Note that this equation is written in moment 
space: only then does the real emission phase--space factorise (see, however, 
\cite{Becher:2006mr}).  Eq.~(\ref{Rad}) describes the structure of the Sudakov 
exponent in terms of a few anomalous dimensions which are functions of the 
running coupling only. This additive structure of the exponent is in one-to-one
correspondence with the phase-space origin of the various corrections: 
collinear logarithms, characterised by momenta of order $m^2 r$, are controlled 
by ${\cal B}(\alpha_s)$; soft (large-angle) logarithms, characterised by 
momenta of order $m^2 r^2$, are controlled by ${\cal D}(\alpha_s)$; 
finally, the overlap between the soft and the collinear regions is governed by 
the cusp anomalous dimension $\gamma_K(\alpha_s)$, which is a universal quantity, the one and only source of double logarithms.

\subsection*{Singularities in multi-leg amplitudes}

\begin{figure}[htb]
\begin{center}
\includegraphics[angle=0,width=8.4cm]{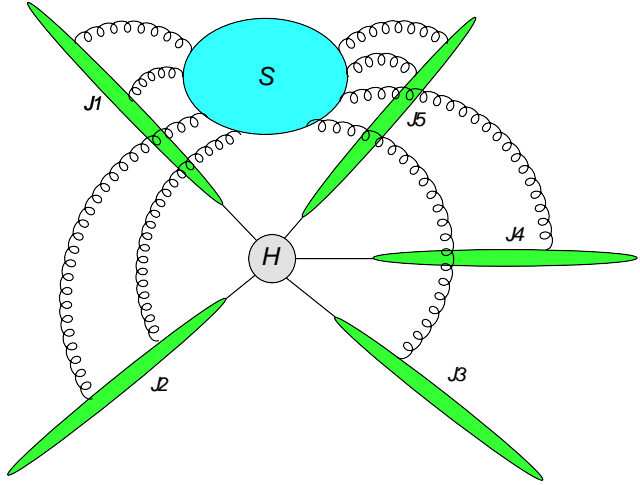}  
\caption{Singular configurations for a fixed-angle multi-parton scattering 
amplitude
\label{fig:multi-jet}}
\end{center}
\end{figure}

The application of Sudakov resummation to hard processes with several 
coloured partons is less developed, and it will become more important for 
LHC physics. The starting point to perform such a resummation is the 
analysis of the singularity structure of the corresponding scattering 
amplitudes (fig.~\ref{fig:multi-jet}) at fixed angles~\cite{Botts:1989kf,
Kidonakis:1997gm}, assuming no strong hierarchy between the various 
kinematic invariants. A priori, upon considering a multi-leg hard process 
with general kinematics, one may expect a complicated singularity structure, 
more intricate than the simple expressions of eqs.~(\ref{Sud}) and (\ref{Rad}).
Yet, the goal remains to understand the singularities to any loop order in terms on 
a small set of anomalous dimensions, which are functions of the coupling only. 

\begin{figure}[htb]
\begin{center}
\begin{tabular}{ccc}
\includegraphics[angle=0,height=2.6cm]{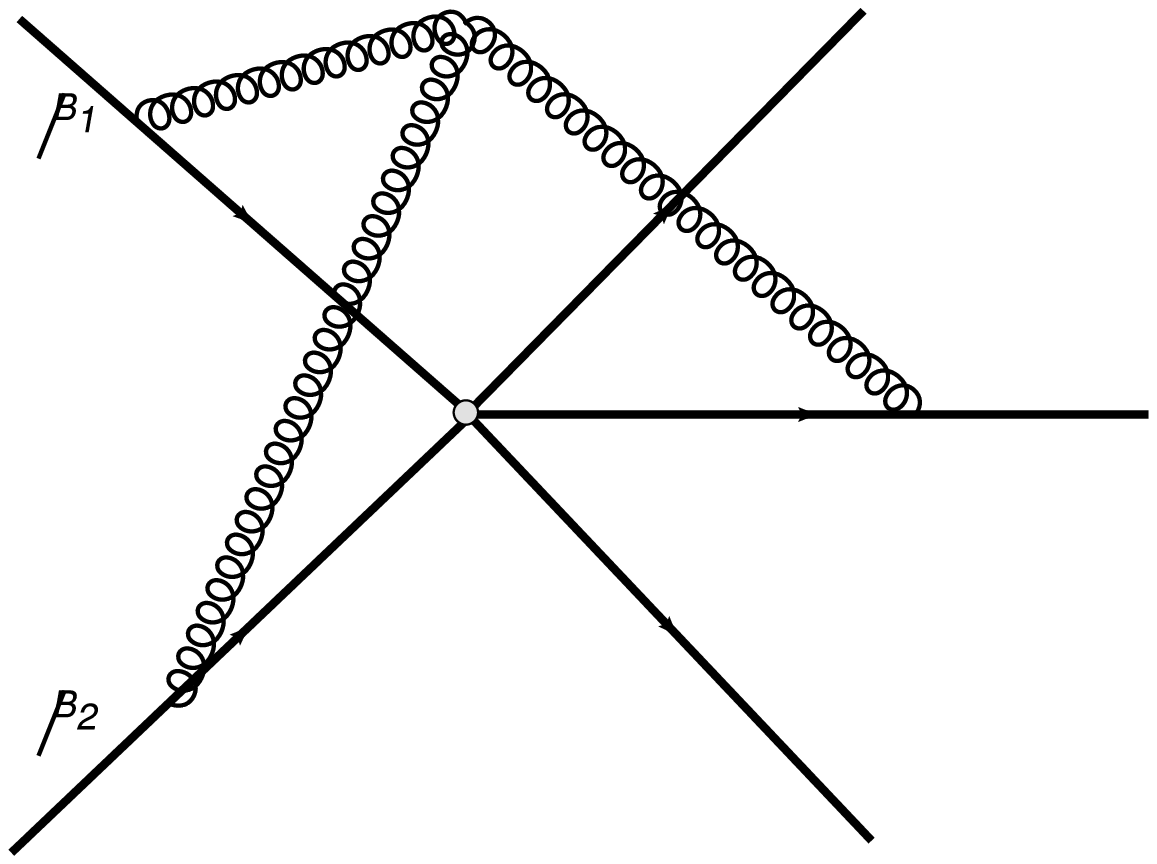}&
\includegraphics[angle=0,height=2.6cm]{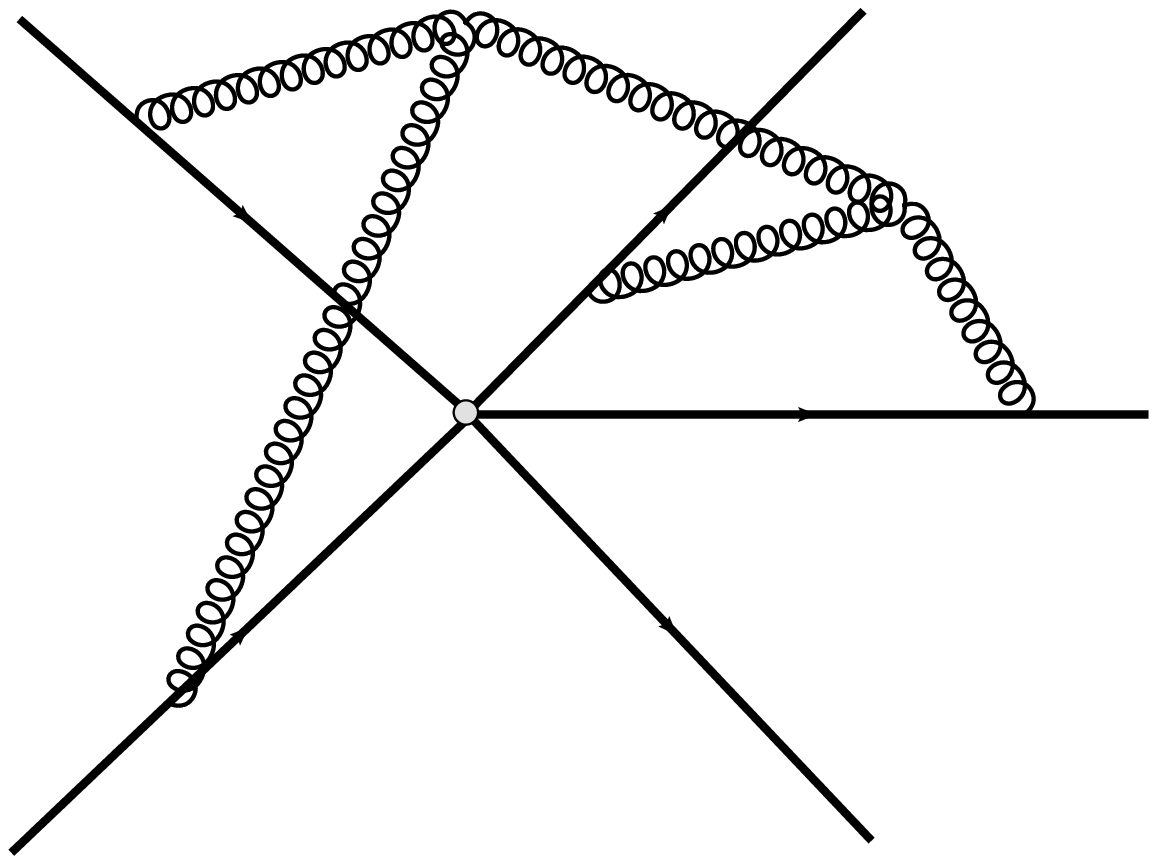}&
\includegraphics[angle=0,height=2.6cm]{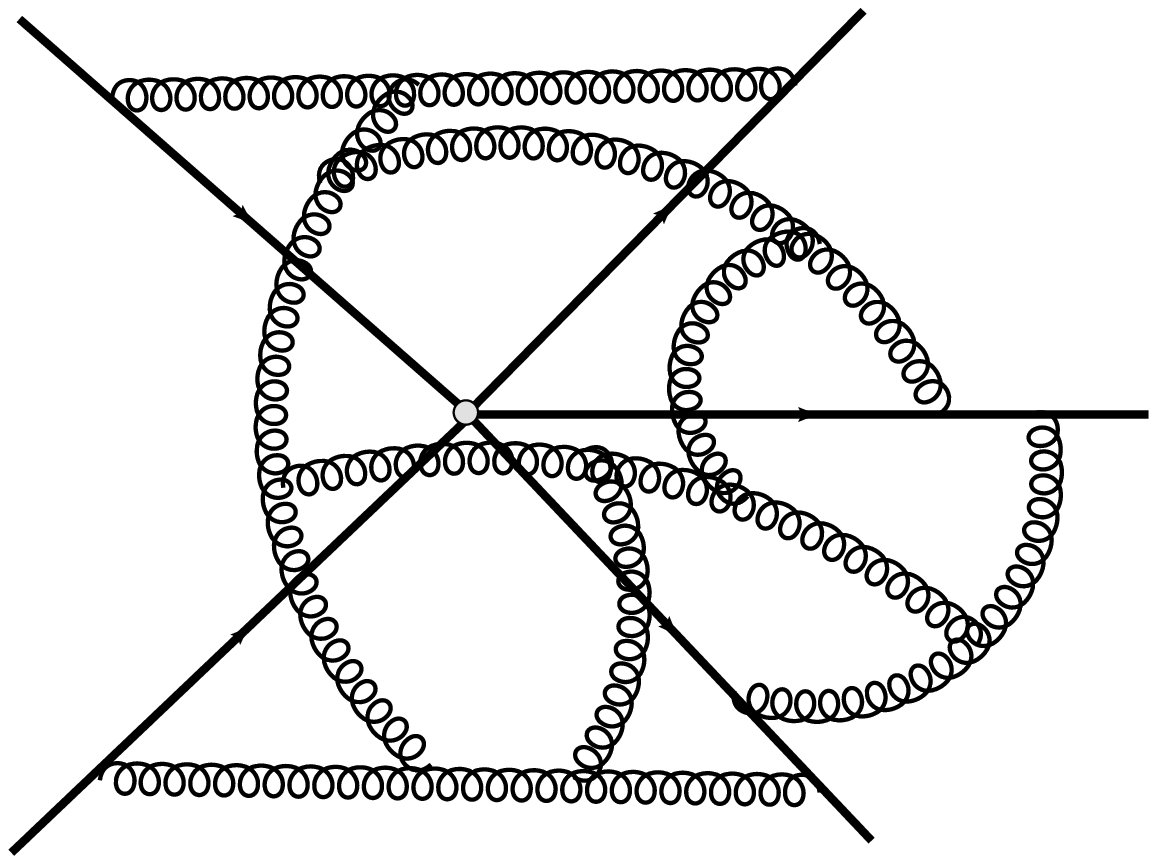}
\end{tabular}
\end{center}
\caption{Gluon `webs' entering the soft function at 2, 3 and 8 loops respectively\label{fig:webs}}
\end{figure}

A further complication arises in a non-Abelian gauge theory, as soon as the 
amplitude involves more than three hard partons: soft gluon interactions induce 
correlations between kinematic and colour degree of freedom. Soft gluons still 
exponentiate, but this exponentiation now involves matrices defined in a given 
colour basis (see below). Resummation formulae such as eqs.~(\ref{Sud}) and 
(\ref{Rad}) may only hold upon diagonalizing these matrices. The size of the
matrices depends only of the colour representations of the external hard partons. 
A priori, however, at each loop order one would expect new colour correlations 
(as suggested by fig.~\ref{fig:webs}), which would require re-diagonalizing the
matrix. 

The theoretical understanding of infrared singularities in multi-leg processes 
has recently taken a significant leap forward. The first step was taken in 
\cite{MertAybat:2006mz}, where a two-loop calculation of the 
infrared singularities in multi-leg amplitudes was first performed. The conclusions
were rather suprising at the time: the colour matrix structure of the soft 
anomalous dimension that controls the singularities at two loops turned out 
to be identical to the one at one loop.  The next step was taken very recently 
in~\cite{Gardi:2009qi} and in parallel in \cite{Becher:2009cu,Becher:2009qa}.
These papers explained the findings of \cite{MertAybat:2006mz} and proposed 
a formula that generalizes this result to all loops. According to this formula, no 
new correlations are introduced by multi-loop webs. Instead,
the correlations generated by soft gluons are always described by a sum over 
two--body interactions between hard partons, and thus 
the matrix structure at any loop order is the same as at one loop. 
We shall present this formula in the next section. This proposal is based on a 
set of all--order constraints (see secs.~\ref{sec:fact_constraints} and 
\ref{sec:solving_the_equations} below) that relate the singularity structure 
in any multi-leg amplitude to the cusp anomalous dimension $\gamma_K
(\alpha_s)$. The derivation of these constraints is based on factorization and 
on the universal properties of soft gluon interactions, which are described in 
secs.~\ref{sec:Eikonal_approx}, \ref {sec:factorization} and \ref{sec:cusp}.
Importantly, in amplitudes with four legs or more the sum-over-dipoles formula is  still an ansatz: although this formula is consistent with all available constraints, 
there may be additional contributions at three loops or beyond, which we can constrain but not exclude at present. This issue is briefly summarized in 
sec.~\ref{sec:beyond}.

\section{The sum-over-dipoles formula\label{sec:ansatz}}

Consider a scattering amplitude ${\cal M} \left(p_i/\mu, \alpha_s (\mu^2),
\epsilon \right)$, involving a fixed number $n$ of hard coloured partons carrying
momenta $p_i$, $i = 1 \, \ldots \, n$, all lightlike, $p_i^2 = 0$, and any 
number of additional non-coloured particles. 
We assume that ultraviolet renormalization has been performed ($\mu$ being the renormalization scale) thus all remaining 
singularities are associated with long-distances, and can be regularized working in $D = 4 - 2 \epsilon$ dimensions, with $\epsilon < 0$.  
The singularities depend on all the kinematic invariants that can be formed out 
of the hard parton momenta, $p_i \cdot p_j$ ($n (n - 1)/2$ invariants for an 
$n-$parton amplitude). We work with general kinematics and assume no special 
hierarchy between these invariants; they must all be large compared to 
$\Lambda_{\rm QCD}^2$, and their ratios are regarded as numbers of order 
unity. Momentum conservation is not imposed between the coloured partons, 
allowing for any recoil momentum to be carried by non-coloured particles in 
both the initial and final states.
Soft and collinear factorization properties guarantee that all infrared singularities 
can be absorbed into an overall multiplicative factor $Z$: one writes formally
\begin{equation}
\label{introducing_Z}
 {\cal M} \left(p_i/\mu, \alpha_s (\mu^2), \epsilon \right)  \, = \,
 Z \left(p_i/\mu_F, \alpha_s (\mu_F^2), \epsilon \right) \,\,
 {\cal H} \left( p_i/\mu, \mu/\mu_F, \alpha_s (\mu^2) \right) \, ,
\end{equation}
where $\cal H$ is finite and can be taken to be independent of 
$\epsilon$. Note that in general, the factorization scale $\mu_F$, at which 
$Z$ is defined, is distinct from the renormalization scale $\mu$. For simplicity 
in the following we choose $\mu_F=\mu$.
Eq.~(\ref{introducing_Z}) should be understood as a matrix multiplication in 
colour space: $\cal H$ is a vector in some colour basis, accounting 
for the hard scattering process, including any loop corrections involving highly 
virtual gluons. These are necessarily finite. $Z$ is a matrix in this space, mixing 
the components of the vector $\cal H$, and accounting for soft and 
collinear singularities. According to the ansatz of Ref.~\cite{Gardi:2009qi} $Z$ 
assumes the form\footnote{Following 
\cite{Catani:1998bh} we keep track of the unitarity phases by writing $- p_i
\cdot p_j = \left| p_i \cdot p_j \right| \, {\rm e}^{{\rm i} \pi 
\lambda_{ij}}$, where $\lambda_{ij} = 1$ if $i$ and $j$ are both initial-state 
partons or are both final--state partons, and $\lambda_{ij} = 0$ otherwise.} 
\begin{align}
\label{Z}  
\begin{split}
 Z \left(p_i/\mu, \alpha_s (\mu^2), \epsilon \right)  = 
 \exp\Bigg\{
 \int_0^{\mu^2}\frac{d\lambda^2}{\lambda^2} \Bigg[
 &\frac18 \,
 \widehat{\gamma}_K\left(\alpha_s(\lambda^2,\epsilon)  \right)  
 \,\sum_{(i,j)} \,
 \ln\left(\frac{2 p_i\cdot p_j\,{\rm e}^{{\rm i} \pi\lambda_{ij} 
 }}{{\lambda^2}}\right) 
 \,  \mathrm{T}_i \cdot   \mathrm{T}_j \,  \\ 
 &  - \,\frac12\, \sum_{i=1}^n \, \gamma_{J_i} \left(\alpha_s(\lambda^2,  
 \epsilon) \right) 
 \Bigg]\Bigg\}\,,
\end{split}
\end{align}
where the notation $\sum_{(i,j)}$ indiactes a sum over all pairs of hard partons forming colour dipoles, where each pair is counted twice $(i,j)$ and $(j,i)$, and~$\mathrm{T}_i \cdot   \mathrm{T}_j\equiv \sum_{a} \mathrm{T}_i^{(a)} \cdot   \mathrm{T}_j^{(a)}$, where $\mathrm{T}_i $ is a generator\footnote{${\rm T}^{(a)}$ should be interpreted as follows: for a final--state quark or an initial--state
antiquark: $t^a_{\alpha \beta}$; for a final--state antiquark or an initial--state quark: $- t^a_{\beta \alpha}$; for a gluon: ${\rm i} \, f_{cab}$.} in the colour representation of parton~$i$. The overall colour charge 
is conserved, 
\begin{align}
\sum_{i = 1}^n \mathrm{T}_i^{(a)} = 0 \, .
\label{T_prop2}
\end{align}
The same ansatz was proposed independently in Ref.~\cite{Becher:2009cu,
Becher:2009qa}.

As originally proposed in \cite{Magnea:1990zb} (see also \cite{Magnea:2000ss}),
singularities in eq.~(\ref{Z}) are generated exclusively through integration over 
the $D$-dimensional running coupling $\alpha_s(\lambda^2,\epsilon)$, which 
obeys the renormalization group equation
\begin{equation}
  \mu \frac{\partial \alpha_s(\mu^2,\epsilon)}{\partial \mu} = \beta(\epsilon, \alpha_s) 
  = - \, 2 \, \epsilon \, \alpha_s
  - \frac{\alpha_s^2}{2 \pi} \, \sum_{n = 0}^\infty b_n \left( 
  \frac{\alpha_s}{\pi} \right)^n~,
\label{beta}
\end{equation}
where $b_0 = (11 C_A - 2 n_f)/3$.
It is easy to verify that the solution to this equation for small coupling and fixed, negative $\epsilon$,  is power suppressed at small scales,
\begin{equation}
 \alpha_s(\lambda^2,\epsilon) = \left(\frac{\lambda^2}{\mu^2}
 \right)^{- \epsilon} \Big[ \alpha_s(\mu^2, \epsilon) + {\cal O}
 \left( \alpha_s^2 \right) \Big] \, ,
\end{equation}
which guarantees convergence of integrals ranging from $\lambda^2 = 0$ to
some fixed scale $\mu^2$, as in eq.~(\ref{Z}). 
Non-trivial higher-loop corrections enter in (\ref{Z}) only through higher--order 
corrections to the anomalous dimensions ${\gamma}_K^{(i)}(\alpha_s)$ and 
$\gamma_{J_i}(\alpha_s)$. The former --- but not the latter --- is assumed 
here to admit Casimir scaling, namely to depend on the colour representation of the parton $i$ only through an overall factor given by the corresponding quadratic Casimir,
\begin{equation}
\label{Casimir_scaling_K}
{\gamma}_K^{(i)} (\alpha_s) = C_i  \, \widehat{\gamma}_K(\alpha_s) \,;
\qquad\quad C_i\equiv \mathrm{T}_i \cdot   \mathrm{T}_i\,.
\end{equation}
$\widehat{\gamma}_K (\alpha_s) = 2 \alpha_s/\pi + {\cal O} (\alpha_s^2)$
is known explicitly to three-loop order based on the calculation by Moch,
Vermaseren and Vogt, \cite{Moch:2004pa,Vogt:2004mw}. Potential 
contributions of higher-order Casimirs at four loops and beyond will be 
briefly discussed in sec.~\ref{sec:beyond}.

The first term in eq.~(\ref{Z}), which is governed by the cusp anomalous 
dimension $\widehat{\gamma}_K$, represents the singularities generated 
by the interaction of large-angle soft gluons described by the ${\cal S}$ 
function in fig.~\ref{fig:multi-jet}. This term is written as a sum over colour 
dipoles formed by any pair of hard partons; it correlates the kinematic 
dependence on the Lorentz invariant $p_i\cdot p_j$ with the corresponding
product of colour generators, $\mathrm{T}_i \cdot   \mathrm{T}_j$. 
This correlation is precisely the one present at the one--loop order. This would 
imply that no new correlations are generated by multi-loop webs such as the 
ones shown in fig.~\ref{fig:webs} --- a highly non-trivial statement, which was not yet tested by direct calculations beyond the two-loop level.

The second term in eq.~(\ref{Z}) represents the interaction of collinear 
gluons. It is governed by the jet anomalous dimension corresponding 
to each of the external partons, quarks or gluons. These anomalous dimensions 
are defined in (\ref{renJ}) below; they depend not only on the colour 
representation of these partons but also on their spin. Their values are known 
to three-loop order, based on the calculation of the quark and gluon form factors 
in Refs.~\cite{Moch:2005id,Moch:2005tm}; the coefficients have been 
conveniently collected in Appendix A of~\cite{Becher:2009qa}.

Eq.~(\ref{Z}) may well be the exact expression for the singularities of any 
on-shell scattering amplitude in massless gauge theories. As already emphasized, 
the simplicity of this result is striking, especially when compared to the lengthy 
and complicated expressions one typically obtains for multi-leg amplitudes.
It is also not what one would naturally expect looking at the diagrams of 
fig.~\ref{fig:webs}. Indeed eq.~(\ref{Z}) requires that some remarkable cancellations take place in these diagrams. It is therefore very interesting to 
see how eq.~(\ref{Z}) emerged out of general considerations. This is the goal 
of the following sections.

\section{Eikonal approximation and rescaling invariance\label{sec:Eikonal_approx}}

The first key ingredient in deriving the constraints on the singularity structure 
is the universal nature of soft gluon interactions, in particular their independence 
on the absolute momentum scale of the hard parton to which they couple. Let us
first explain the origin of this property and then analyse its consequences in the
context of the factorization of an on-shell amplitude. 

\begin{figure}[htb]
\begin{center}
\includegraphics[angle=0,width=4.4cm]{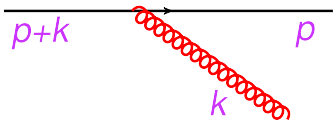}  
\caption{Gluon bremsstrahlung off an outgoing quark. The final--state quark is on shell, $p^2=m^2$.\label{fig:brem}}
\end{center}
\end{figure}

Consider, for example, soft gluon radiation off a hard quark, as shown in 
fig.~\ref{fig:brem}. We assume that after this emission the quark is on shell,
so that $p^2 = m^2$ (where $m^2$ can vanish, but this is not necessary 
for the argument that follows). Applying the ordinary Feynman rules one 
obviously obtains a result that depends on the radiating particle spin and 
momentum --- the first expression in eq.~(\ref{eikonal_example}). Considering 
instead the limit where the gluon is soft ($k\to 0$) one obtains a much simpler result,
\begin{equation}
\label{eikonal_example}
\bar{u}(p)\left(-{\rm i}g_sT^{(a)}\gamma^{\mu}\right) \,\frac{{\rm i}(\psl+\ksl+m)}{(p+k)^2-m^2+{\rm i}\varepsilon}\,
\qquad 
\begin{array}{c}
\\
{\Longrightarrow\,\,}\\
{\hspace{-1.5mm} k\to 0}
\end{array}
\qquad
\bar{u}(p) g_s T^{(a)} \frac{p^{\mu}}{p\cdot k+{\rm i}\varepsilon} \, ,
\end{equation}
which depends only on the colour charge and direction $\beta$ of the quark 
momentum
\begin{equation}
 g_s T^{(a)} \frac{{p^{\mu}}}{{ p}\cdot k+{\rm i}\varepsilon} = g_s T^{(a)} \frac{{ \beta^{\mu}}}{{\beta}\cdot k+{\rm i}\varepsilon}\,.
\end{equation}
Equivalently, we observe that the resulting ``eikonal'' Feynman rules are invariant with respect to rescaling of the quark velocity, $\beta \to  \kappa  \beta$,
a symmetry property that will be central to our discussion in what follows. 
Finally, we note that the eikonal approximation is conveniently formulated by
replacing the dynamical hard partons, which provide the source for the radiation, 
with Wilson lines along their classical trajectories,
\begin{equation}
\label{Wilson_line}
  \Phi_{\beta} (0, -\infty) =
  P \exp \left[\, {\rm i} g_s \int_{-\infty}^{0} d \lambda \,
  \beta \cdot A(\lambda \beta) \, \right]~.
\end{equation}
Here rescaling invariance is inherent: it is realised through reparametrization of the integral along the path.

\section{Factorization\label{sec:factorization}} 

The second key element is the factorization of soft and collinear singularities 
in the amplitude, illustrated in fig.~\ref{fig:multi-jet}. Following \cite{Sen:1982bt,
Kidonakis:1998nf,Sterman:2002qn,MertAybat:2006mz,Dixon:2008gr}, we write
\begin{eqnarray}
\label{facamp}
  {\cal M}_{L} \left(p_i/\mu, \alpha_s (\mu^2),
  \epsilon \right) & = & \sum_K
  {\cal S}_{L K} \left(\beta_i \cdot \beta_j, \alpha_s (\mu^2), \epsilon 
  \right) \,  H_{K} \left( \frac{2 p_i \cdot p_j}{\mu^2},
  \frac{(2 p_i \cdot n_i)^2}{n_i^2 \mu^2}, \alpha_s (\mu^2) \right)
  \nonumber \\ && \times
  \prod_{i = 1}^n \frac{{\displaystyle J_i 
  \left(\frac{(2 p_i \cdot n_i)^2}{n_i^2 \mu^2},
  \alpha_s (\mu^2), \epsilon \right)}}{{\displaystyle {\cal J}_i 
  \left(\frac{2 (\beta_i \cdot n_i)^2}{n_i^2}, \alpha_s (\mu^2), 
  \epsilon \right)}  \,} \,\, .
\end{eqnarray}
Here the hard function $H_{K}$ and the amplitude ${\cal M}_{L}$ 
are vectors in the space of available color configurations; the soft function 
${\cal S}_{L K}$ is a matrix in this space, while the jet functions 
$J_i$ and  ${\cal J}_i$ do not carry any colour index. The soft matrix 
${\cal S}$ and the jet functions $J$ and ${\cal J}$ contain all soft and collinear singularities of the amplitude, while the hard functions $H_K$ 
can be taken to be independent of $\epsilon$. Each of the functions appearing 
in eq.~(\ref{facamp}) is separately gauge invariant and admits operator 
definitions that are given below. 

The soft and jet functions involve semi-infinite Wilson lines, defined in 
(\ref{Wilson_line}). The `partonic jet' function (for, say, an outgoing 
quark with momentum $p$) is defined by
\begin{equation}
  \overline{u}(p)  \, J \left( \frac{(2p \cdot n)^2}{n^2 \mu^2}, 
  \alpha_s(\mu^2), \epsilon \right) \, = \, \langle p \, | \, 
  \overline{\psi} (0) \, \Phi_n (0, - \infty) \,  | 0 \rangle \, .
\label{Jdef}
\end{equation}
The function $J$ represents a transition amplitude connecting the 
vacuum and a one-particle state. The eikonal line $\Phi_n$ simulates
interactions with partons moving in different directions: the direction
$n^\mu$ is arbitrary, but off the light-cone, in order to avoid spurious 
collinear singularities. 

The factorization formula (\ref{facamp}) also involves the 
eikonal approximation to the partonic jet $J$, which we call the `eikonal 
jet'. It is defined by
\begin{equation}
  {\cal J} \left( \frac{2 (\beta \cdot n)^2}{n^2}, \alpha_s(\mu^2), \epsilon 
  \right)  \, = \, \langle 0 | \, \Phi_{\beta}(\infty, 0) \, 
  \Phi_{n} (0, - \infty) \, | 0 \rangle~,
\label{calJdef}
\end{equation}
where the velocity vector $\beta_i$ of each jet is related to the corresponding 
momentum $p_i$ by $p_i =  \beta_i Q_0/\sqrt{2}$, with $Q_0$ a hard scale
such that $p_i \cdot p_j/ Q_0^2$ is of order one for all $i,j$.

Both the partonic jet (\ref{Jdef}) and the eikonal jet (\ref{calJdef}) have 
soft divergences, as well as collinear divergences associated to their 
light-like leg; thus, they display double poles order by order in perturbation 
theory. The double poles are however the same, since in the 
soft region ${\cal J}$ correctly approximates $J$: singular contributions 
to the two functions differ only by hard collinear radiation. 

The final ingredient in (\ref{facamp}) is the soft matrix, which we define by 
taking the eikonal approximation for all gluon exchanges: since soft gluons
do not resolve the details of the hard interaction nor the internal structure of 
the jets, they couple effectively to Wilson lines in the colour representations 
of the corresponding hard external partons. Such exchanges mix the colour
components of the amplitude, forming a matrix in colour space. Choosing a 
basis of independent tensors $c_L$ in color space, we write
\begin{equation}
\sum_L \left( c_L \right)_{\{\alpha_k\}} {\cal S}_{L K} \left(\beta_i 
\cdot \beta_j, \alpha_s(\mu^2), \epsilon \right) = \sum_{\{\eta_k\}} \, \, \langle 0 |  \,
\prod_{i = 1}^n \Big[ \Phi_{\beta_i} (\infty, 0)_{\alpha_k, \eta_k} 
\Big]  \, | 0 \rangle  \, \left( c_K \right)_{\{\eta_k\}} \, .
\label{softcorr}
\end{equation}
Note that in eq.~(\ref{softcorr}) we keep all Wilson lines strictly on the light-cone 
($p_i^2 = 0$ and thus $\beta_i^2 = 0$). Therefore, the soft matrix ${\cal 
S}_{L K}$ displays not only single poles corresponding to large-angle soft gluons, 
but also double poles associated with overlapping soft and collinear singularities. 
Recall that the jet functions $J_i$ also include the regions of overlapping soft and 
collinear singularities. It is for this reason that in the factorization formula, 
eq.~(\ref{facamp}), we have divided each partonic jet $J_i$ jet by ${\cal J}_i$, 
thus removing from $J_i$ its eikonal part, which is already accounted for in 
${\cal S}_{L K}$.  One observes then that the ratios $J_i/{\cal J}_i$ are free 
of soft singularities: they contain only single collinear poles at each order in 
perturbation theory. Similarly, the `reduced' soft matrix
\begin{equation}
  \overline{{\cal S}}_{L K} \left(\rho_{i j},\alpha_s(\mu^2), \epsilon \right) = 
  \frac{{\cal S}_{L K} \left(\beta_i 
  \cdot \beta_j, \alpha_s(\mu^2), \epsilon \right)}{\displaystyle \prod_{i = 1}^n 
  {\cal J}_i \left(\frac{2(\beta_i \cdot 
  n_i)^2}{n_i^2}, \alpha_s(\mu^2), \epsilon \right)}
\label{reduS}
\end{equation}
where
\begin{equation}
\rho_{ij} \equiv \frac{  \, (\beta_i \cdot \beta_j \,{\rm e}^{{\rm i} \,
\pi\lambda_{ij}})^2 } {\displaystyle \frac{2(\beta_i \cdot n_i)^2}{n_i^2} 
\frac{2(\beta_j \cdot n_j)^2}{n_j^2}} \,  .
\label{rhoij}
\end{equation}
is free of collinear poles, and contains only infrared singularities originating 
from soft gluon radiation at large angles with respect to all external legs. 
The kinematic dependence of $\overline{{\cal S}}$ on $\rho_{ij}$ will be 
explained below.

\section{The cusp anomaly and the jet functions\label{sec:cusp}} 

The third and final ingredient necessary for deriving the constraints on the singularities of on-shell amplitudes is the cusp anomaly. In the following we will recall the definition of the cusp anomalous dimension and explain its role in governing the kinematic dependence of Wilson--line operators. This will allow us
first to understand the structure of the eikonal jet function ${\cal J}$ to all 
orders in perturbation theory, and eventually to constrain the soft function 
${\cal S}$.

To this end, let us recall some general properties of operators that are 
composed of semi-infinite Wilson lines. 
The first observation is that all radiative corrections to such operators vanish 
identically in dimensional regularization, since the corresponding integrals involve 
no scale. This trivial result however involves cancellations between ultraviolet
and infrared singularities; therefore, upon renormalization, ${\cal J}$ becomes 
non-trivial: the contribution of each graph equals minus the corresponding 
ultraviolet counterterm. As a consequence, using a minimal subtraction scheme,
the result for ${\cal J}$ (or for ${\cal S}$) at each order in $\alpha_s$ is a sum 
of poles in $\epsilon$, without any non-negative powers.  

Let us now briefly recall the renormalization properties of Wilson loops with 
cusps \cite{Polyakov:1980ca,Dotsenko:1979wb,Brandt:1981kf,
Korchemsky:1985xj,Ivanov:1985np,Korchemsky:1987wg,Korchemsky:1988hd,
Korchemsky:1988si,Korchemskaya:1994qp}.  Consider first an operator
\begin{equation}
  W \left( \gamma_{12}, \alpha_s(\mu^2), \epsilon 
  \right)  \, = \, \langle 0 | \, \Phi_{n_1}(\infty, 0) \, 
  \Phi_{n_2} (0, - \infty) \, | 0 \rangle~,
\label{calWdef}
\end{equation}
involving two semi-infinite rays, both off the lightcone ($n_1^2,\, n_2^2 
\neq 0$), which join at the origin to form a cusp with (Minkowski) angle $\gamma_{12}$, where
\begin{equation}
\label{gamma12}
\cosh\left(\gamma_{12}\right)= \frac{(n_1\cdot n_2)}{\sqrt{n_1^2}\,\sqrt{n_2^2}} \, .
\end{equation}
The contour closes at infinity and it is smooth everywhere except at the 
origin. Ref.~\cite{Brandt:1981kf} has shown that the presence of the cusp 
along the contour introduces an ultraviolet singularity which can be removed 
by a multiplicative renormalization constant, implying that
\begin{align}
\label{rge_W}
 \frac{d \ln W }{d\ln\mu} & \equiv - \, \Gamma_{\rm cusp} 
 (\gamma_{12}, \alpha_s (\mu^2)) \, = \, - \, C_i \,
 \frac{\alpha_s(\mu^2)}{\pi} \,\Big[\gamma_{12}
 \coth(\gamma_{12}) - 1 \Big] \, + {\cal O} \left( \alpha_s^2 \right) \, ,
\end{align}
where $C_i = C_A$ or $C_F$ depending on the representation of the Wilson 
lines. Considering the limit where $n_1$ or $n_2$ is near the lightcone one 
finds that 
\begin{equation}
\gamma_{12}\simeq \ln \left(\frac{2n_1\cdot n_2}{{\sqrt{n_1^2}\,\sqrt{n_2^2}}}\right)\,\gg\, 1 \, ,
\end{equation}
and 
\begin{align}
\label{rge_W1}
\frac{d \ln W\left( \gamma_{12}, \alpha_s(\mu^2), \epsilon 
  \right) }{d\ln\mu} &=-\frac{\gamma_K^{(i)}(\alpha_s(\mu^2))}{2}
 \,\ln \left(\frac{2n_1\cdot n_2}{\sqrt{n_1^2}\,\sqrt{n_2^2}}\right)+
 {\cal O}(1) \, .
\end{align}
It is clear that in the strictly light-like limit, the {\it r.h.s.} of (\ref{rge_W1}) 
and (\ref{rge_W}) become singular. This is a collinear singularity, appearing 
on top of the ultraviolet singularity already present in $W$ owing to the cusp.
Therefore, if we consider directly the renormalization of the analogue of 
$W$ with one of the rays being strictly lightlike, say $n_2^2 = 0$ --- precisely 
the case of ${\cal J}$ in eq.~(\ref{calJdef}) --- we expect a singular anomalous
dimension. Indeed, in dimensional regularization the renormalization group 
equation of ${\cal J}$ takes the form~\cite{Gardi:2009qi}
\begin{eqnarray}
\label{rencalJ}
  && \hspace{1cm} \mu \frac{d}{d \mu} 
  \ln {\cal J}_i \left( \frac{2 (\beta \cdot n)^2}{n^2}, \alpha_s(\mu^2), 
  \epsilon \right) \, \equiv \, - \, \gamma_{{\cal J}_i} \left(\frac{2 (\beta 
  \cdot n)^2}{n^2}, \alpha_s(\mu^2), \epsilon \right) \\
 && \hspace*{10pt}
 = \frac12 \, \delta_{{\cal J}_i} \left(\alpha_s(\mu^2) \right) \, 
 - \frac14 \gamma_K^{(i)} \left( \alpha_s(\mu^2) \right) \, \ln 
 \left(\frac{2 (\beta \cdot n)^2}{n^2}\right) \,
 - \frac14 \int_0^{\mu^2} \frac{d \xi^2}{\xi^2} \, \gamma_K^{(i)} 
 \left( \alpha_s (\xi^2,\epsilon) \right) \, , 
 \nonumber
\end{eqnarray}
where the third term is singular, ${\cal O}(1/\epsilon)$.
Having seen the origin of this singularity in (\ref{rge_W1}), it is not surprising  
that the dependence of $\gamma_{\cal J}$ on the 
kinematic variable ${2 (\beta \cdot n)^2}/{n^2}$ --- the second term in 
(\ref{rencalJ}) --- is governed by the same anomalous dimension, 
$\gamma_K^{(i)} (\alpha_s)$, that governs the singularity of 
$\gamma_{{\cal J}_i}$ --- the third term in (\ref{rencalJ}). This relation 
between kinematic dependence and singular terms, which we have now 
observed in $\gamma_{{\cal J}_i}$, is a general property of this class of 
operators which will be essential for what follows. 

To understand it from a different angle, let us now have another look at 
eqs.~(\ref{rge_W}) and~(\ref{rencalJ}) considering the symmetry property of the eikonal Feynman rules (\ref{eikonal_example}) under rescaling of the eikonal velocity vectors. 
Clearly eq.~(\ref{rge_W}) is consistent with this symmetry: any function of 
$\gamma_{12}$ defined in~(\ref{gamma12}) would be. In contrast, in the 
strictly lightlike case of~(\ref{rencalJ}), there is no kinematic variable that 
could be consistent with this symmetry.  ${\cal J}$ can only depend on 
${2 (\beta \cdot n)^2}/{n^2}$, as indeed can be confirmed by an explicit
calculation, and therefore it breaks the rescaling symmetry: it depends 
explicitly on the normalization of~$\beta$. Note that rescaling of the vector 
$n_{\mu}$, which is not light-like, remains a symmetry.  

Solving (\ref{rencalJ}) we obtain a closed form expression for the eikonal 
jet~\cite{Gardi:2009qi}, in terms of anomalous dimensions which depend
just on the coupling,
\begin{eqnarray}
 {\cal J}_i \left(\frac{2(\beta_i\cdot n_i)^2}{n_i^2}, \alpha_s (\mu^2), \epsilon   
 \right)  =  \exp \Bigg\{ &&
 \frac12 \int_{0}^{\mu^2} \frac{d \lambda^2}{\lambda^2}
 \bigg[ \frac12 \delta_{{\cal J}_i} \Big( \alpha_s( \lambda^2, \epsilon ) 
 \Big) \nonumber \\
 \label{fincalJ} 
 && \, - \, \frac14 \gamma_K^{(i)} \Big( \alpha_s(\lambda^2, \epsilon)
 \Big) \, \ln \left(\frac{2(\beta_i\cdot n_i)^2\,\mu^2}{n_i^2\lambda^2} 
 \right) \bigg] \Bigg\} \, .
\end{eqnarray}
We see that the entire kinematic dependence of ${\cal J}$ is associated 
with the breaking of the rescaling symmetry with respect to the lightlike
direction $\beta_{\mu}$; it is directly related to presence of double poles 
in ${\cal J}$, and it is governed by the cusp anomalous dimension, 
$\gamma_K^{(i)} (\alpha_s)$. In the following, we will show how this 
observation, made in \cite{Gardi:2009qi}, allowed us to constrain the 
kinematic dependence of the soft function ${\cal S}$.

Before turning to the soft function, let us quote the equivalent expression 
for the partonic jet $J_i$, which will be of use in the following. 
The partonic jet has an infrared singularity structure similar to the eikonal 
jet, however it has a \emph{finite} ultraviolet anomalous dimension, the one 
we have encountered in (\ref{Z}),
\begin{equation}
  \mu \frac{d}{d \mu} 
  \ln J_i \left( \frac{(2 p_i \cdot n_i)^2}{n_i^2 \mu^2}, \alpha_s(\mu^2),  
  \epsilon \right) 
  \equiv - \, \gamma_{J_i} (\alpha_s(\mu^2)) \,.
\label{renJ}
\end{equation}
As a consequence, the partonic jet function can be written as
\begin{eqnarray}
\label{J_explicit}
 && \hspace{-2mm}
 J_i \left(\frac{(2p_i\cdot n_i)^2}{n_i^2\mu^2}, \alpha_s(\mu^2), \epsilon 
 \right) = \, H_{J_i} \hspace{-1mm} \left( \alpha_s \left(
 \frac{(2 p_i\cdot n_i)^2}{n_i^2}\right), \epsilon\right) 
 \exp \Bigg\{ \! - \frac12 \int_{0}^{\mu^2} \!\!
 \frac{d\lambda^2}{\lambda^2} 
 \gamma_{J_i} \left( \alpha_s(\lambda^2,\epsilon) \right) \nonumber \\
 && \hspace{8mm} 
 + \, \frac{\mathrm{T}_i \cdot   \mathrm{T}_i}{2} \, \hspace{-1pt}
 \int_0^{{\frac{(2 p_i\cdot n_i)^2}{n_i^2}}} 
 \frac{d \lambda^2}{\lambda^2} 
 \Bigg[ -\frac14 \widehat{\gamma}_K \left( \alpha_s( \lambda^2, 
 \epsilon) \right)
 \ln \left({\displaystyle\frac{(2 p_i\cdot n_i)^2}{\lambda^2 \,n_i^2}} 
 \right) +   \frac12 \widehat{\delta}_{{\overline{\cal S}}}(
 \alpha_s(\lambda^2,\epsilon)) \Bigg] \Bigg\} \, , 
\end{eqnarray}
where $H_J$ is a finite coefficient function, independent of~$\mu^2$.
Eq.~(\ref{J_explicit}) displays the fact that in addition to the collinear 
singularities generated by~$\gamma_{J_i}$ the jet function~(\ref{Jdef}) 
involves soft~(eikonal) singularities; these are summarized by the second 
line of~(\ref{J_explicit}). Beyond the $\gamma_K^{(i)}$ terms, one finds
single pole terms governed by ${\delta}^{(i)}_{{\overline{\cal S}}}$. This 
function, which is defined in eqs. (4.7) and (4.9) in~\cite{Gardi:2009qi}, does 
not depend on the spin of parton $i$ and it has a maximally non-Abelian 
structure. For simplicity, we further assumed here that ${\delta}^{(i)}_{
{\overline{\cal S}}}$ admits Casimir scaling, ${\delta}^{(i)}_{{\overline{\cal S}}}
= \mathrm{T}_i \cdot   \mathrm{T}_i \,\widehat{\delta}_{{\overline{\cal S}}}$,
although this may not hold beyond three loops (and it would not be important 
in what follows). In contrast, $\gamma_{J_i}$, which governs the collinear
singularities, does depend on the spin of parton $i$ (it differs for quarks
and for gluons, see Appendix A of~\cite{Becher:2009qa}) and it is not 
maximally non-Abelian. As anticipated, the double poles in (\ref{fincalJ}) and 
(\ref{J_explicit}) are the same, while single poles differ.    

\section{Derivation of the constraints on soft singularities
\label{sec:fact_constraints}}

We are finally in a position to derive the promised constraints on soft singularities.
We will show, in particular, that the relation established above, considering the 
case of the eikonal jet, between kinematic dependence of single pole terms and
the cusp anomalous dimension, generalises to soft singularities in multi-leg 
amplitudes. These singularities are described by the function ${\cal S}$ in 
(\ref{facamp}). ${\cal S}$ is defined by (\ref{softcorr}) and it obeys a matrix
evolution equation of the form
\begin{equation}
 \mu  \frac{d}{d \mu} {\cal S}_{I K} \left( \beta_i \cdot 
 \beta_j, \alpha_s, \epsilon \right) = - \, \sum_{J}
 \Gamma^{{\cal S}}_{I J} \left( \beta_i \cdot \beta_j, \alpha_s, 
 \epsilon \right) \,\, {\cal S}_{J K} \left( \beta_i \cdot \beta_j, 
 \alpha_s, \epsilon \right) \, .
\label{renS}
\end{equation}
The soft anomalous dimension matrix $\Gamma^{{\cal S}}$ depends on all the
kinematic invariants in the process, and it is a priori a very complicated object. 
It encapsulates the correlation between colour and kinematic degrees of freedom, which may be of increasing complexity as one considers higher loop corrections (fig.~\ref{fig:webs}). 
 
We recall that in ${\cal S}$ all the Wilson lines are lightlike, $\beta_i^2=0$, $\forall i$. Therefore, similarly 
to ${\cal J}$, we expect this function to break the rescaling invariance with
respect to each of the velocities~$\beta_i$. This was already taken into account 
in assigning the arguments in ${\cal S}$ and in $\Gamma^{{\cal S}}$: these
functions depend on the set of Lorentz invariants $\beta_i\cdot \beta_j$, and
thus violate the rescaling symmetry. We will be able to constrain 
$\Gamma^{{\cal S}}$ -- and thus ${\cal S}$ --- because we know 
exactly how this symmetry is violated. 

The key point is that the amplitude ${\cal M}$ itself cannot depend on the
normalization one chooses for the velocities appearing in eikonal functions. 
Thus, in the factorization formula, eq.~(\ref{facamp}), this dependence must
cancel out. This cancellation can only involve the eikonal functions ${\cal S}$ 
and ${\cal J}_i$, not the partonic jet or the hard function, which depend 
directly on the dimensionful kinematic variables $p_i$.  The form of the
factorization formula (\ref{facamp}) implies in fact that the cancellation of 
any rescaling violation must occur within the reduced soft function 
$\overline{\cal S}$, defined in (\ref{reduS}). This is intuitively clear: we saw 
that rescaling violation is intimately related to the presence of double poles,  and
that both are governed by the cusp anomalous dimension $\gamma_K^{(i)}$. 
The soft function ${\cal S}$, much like the eikonal jets, is defined with light-like
Wilson lines, thus including regions of overlapping ultraviolet and collinear
singularities, which are the origin of double poles as well as rescaling violation 
at the single pole level. Upon dividing ${\cal S}$ by the product of all 
eikonal jets, as done in (\ref{reduS}), these regions are removed, yielding 
$\overline{\cal S}$, which describes large--angle soft singularities, and is 
entirely free of double poles and of the associated violation of rescaling 
symmetry at the single pole level. Given the kinematic dependence of 
${\cal S}$ and ${\cal J}_i$, and the expected recovery of the symmetry 
$\beta_i \to \kappa_i \beta_i$, $\forall i$, we deduce that $\overline{\cal S}$
can only depend on the variables $\rho_{ij}$, defined in (\ref{rhoij}). 

To proceed, it is useful to consider the renormalization group equation for the
reduced soft function. In analogy with (\ref{renS}) we have
\begin{equation}
\mu  \frac{d}{d \mu} \overline{{\cal S}}_{I K}
\left( \rho_{i j}, \alpha_s, \epsilon \right) = - \,\sum_{J} 
\Gamma^{\overline{{\cal S}}}_{I J} \left( \rho_{i j}, 
\alpha_s \right) \,\, \overline{{\cal S}}_{J K}
\left( \rho_{i j}, \alpha_s, \epsilon \right) \,.
\label{rencalS} 
\end{equation}
In contrast to $\Gamma^{{\cal S}}$, the anomalous dimension of the reduced 
soft function, $\Gamma^{\overline{{\cal S}}}$, is finite ($\overline{{\cal S}}$ 
itself has only single poles) and invariant under rescalings, as reflected in the 
fact that it must depend on $\beta_i$ though $\rho_{ij}$ only.  Using the
definition of the reduced soft function in eq.~(\ref{reduS}) we can directly 
relate the anomalous dimension $\Gamma^{\overline{{\cal S}}}$ to 
$\Gamma^{{\cal S}}$ and to the anomalous dimension of the eikonal jets, 
$\gamma_{{\cal J}_i}$ of eq.~(\ref{rencalJ}). We obtain
\begin{eqnarray}
  \Gamma^{\overline{{\cal S}}}_{I J} \left(\rho_{i j}, \alpha_s \right) \,
  & = & \, \Gamma^{{\cal S}}_{I J} \left( \beta_i \cdot \beta_j, \alpha_s, 
  \epsilon \right) - \delta_{I J} \sum_{k = 1}^n \gamma_{{\cal J}_{k}} 
  \left(\frac{2(\beta_k \cdot n_k)^2}{n_k^2}, 
  \alpha_s, \epsilon \right) \nonumber \\
  & = &  \, \Gamma^{{\cal S}}_{I J} \left( \beta_i \cdot \beta_j, 
  \alpha_s, \epsilon\right) - \delta_{I J} \sum_{k = 1}^n
  \bigg[-\frac12 \delta_{{\cal J}_{k}}\left( \alpha_s
  \right)  \label{constraints_integral_form} \\ 
  && + \, \frac14 \, \gamma_K^{(k)} \left( \alpha_s 
  \right) \, \ln \left(\frac{2(\beta_k \cdot n_k)^2}{n_k^2}\right) \, 
  + \, \frac14 \int_0^{\mu^2} \frac{d \xi^2}{\xi^2} \gamma_K^{(k)} 
  \Big( \alpha_s (\xi^2,\epsilon) \Big)\bigg] \, . \nonumber 
\end{eqnarray}
This equation implies highly non-trivial constraints on soft singularities. It
tells us precisely how the double poles and the rescaling violation of the single 
poles in $\Gamma^{{\cal S}}$ cancel out. In particular, observing that the jet
terms $\gamma_{{\cal J}_{k}}$  are diagonal in colour space (they are
proportional to the identity matrix), we deduce that

\begin{itemize}
\item{} off diagonal terms in $\Gamma^{{\cal S}}$ must be finite, and 
must depend only on \emph{conformal cross ratios},  
\begin{equation}
\label{rhoijkl}
 \rho_{ijkl} \equiv \frac{(\beta_i \cdot \beta_j) (\beta_k 
 \cdot \beta_l)}{(\beta_i \cdot \beta_k) (\beta_j \cdot \beta_l)}
 = \left(\frac{\rho_{i j} \, \rho_{k l}}{\rho_{i k} \, \rho_{j l}} 
 \right)^{1/2} {\rm e}^{-{\rm i}\pi(\lambda_{ij} + \lambda_{kl} - 
 \lambda_{ik} -\lambda_{jl})}
\end{equation}
which can be interchangeably expressed in terms of $\beta_i \cdot \beta_j$
(the arguments of $\Gamma^{{\cal S}}$), or in terms of $\rho_{ij}$ (the
arguments of $\Gamma^{\overline{\cal S}}$);

\item{} diagonal terms in $\Gamma^{{\cal S}}$ have a singularity
determined by $\gamma_K$, according to
\begin{equation}
\Gamma^{{\cal S}}_{I J} \left( \beta_i \cdot \beta_j, \alpha_s, \epsilon
\right) = \delta_{I J} \sum_{k = 1}^n 
\, \frac14 \int_0^{\mu^2} \frac{d \xi^2}{\xi^2} \gamma_K^{(k)} \Big(
\alpha_s (\xi^2,\epsilon) \Big) + {\cal O}(\epsilon^0)
\end{equation}
and must contain finite terms depending on $\beta_i \cdot \beta_j$ in a way tailored to combine with the $(\beta_i\cdot n_i)^2 /n_i^2$ 
dependence of the various jet functions to generate $\rho_{ij}$.
\end{itemize}

The constraints on the structure of the anomalous dimension 
$\Gamma^{\overline{\cal S}}$ can be compactly expressed by taking 
a logarithmic derivative of eq.~(\ref{constraints_integral_form}) with 
respect to $(\beta_i \cdot n_i)^2/{n_i^2}$. On the {\it l.h.s} one uses the chain 
rule: for any function $F$ which depends on $(\beta_i \cdot n_i)^2/{n_i^2}$ only through the combinations $\rho_{ij}$ of (\ref{rhoij}), one has
\begin{equation}
\frac{\partial}{\partial \ln \left((\beta_i \cdot n_i)^2/{n_i^2}\right)} F \left( \rho_{i j} \right) = 
- \sum_{j \neq i} \frac{\partial}{\partial \ln \rho_{i j}} 
F \left(\rho_{i j} \right) \, .
\end{equation} 
On the {\it r.h.s} of eq.~(\ref{constraints_integral_form}), the derivative 
with respect to $(\beta_i \cdot n_i)^2/{n_i^2}$ acts only on the corresponding $\gamma_{{\cal J}_{i}}$
term. The resulting equations are 
\begin{equation}
\label{constraints}
\sum_{j \neq i} \frac{\partial}{\partial \ln(\rho_{i j})} \,
\Gamma^{{\overline{\cal S}}}_{IJ} \left( 
\rho_{i j}, \alpha_s \right)  =  \frac{1}{4} \, \gamma_K^{(i)} 
\left( \alpha_s \right) \,\delta_{IJ}\,, \qquad \qquad \forall i,\,I,J \, .
\end{equation}
Thus, there are $n$ constraints for an $n$ legged amplitude, each of which 
is a matrix equation (holding for each matrix element $(I,J)$). This set of
constraints holds in any colour basis, and to all orders in perturbation theory. 
Its most intriguing aspect is that it correlates the kinematic dependence of 
the (reduced) soft matrix with its dependence on the colour degrees of 
freedom: the {\it l.h.s} in (\ref{constraints}) is a sum of non-diagonal matrices
in colour space, while the {\it r.h.s} is proportional to the identity matrix.
 
\section{Solving the equations\label{sec:solving_the_equations}}
 
Given $n$ independent equations and $n (n - 1)/2$ kinematic variables it 
is clear at the outset that eq.~(\ref{constraints}) alone is not sufficient to 
uniquely fix the kinematic dependence of $\Gamma^{{\overline{\cal S}}}$ 
in the multi-leg case.  For $n = 2,3$ eq.~(\ref{constraints}) does have a 
unique solution (see sec.~4 and Appendix A in~\cite{Gardi:2009qi}). This is 
already an important step, extending previously known results for the singularity 
structure to all loops. For $n\geq 4$ partons, however, the number of kinematic
variables exceeds the number of equations, and additional constraints will be 
needed. Nevertheless, we will see that a minimal solution, consistent with all 
information known to date, naturally emerges out of eq.~(\ref{constraints}).
   
Considering eq. (\ref{constraints}), we note that $\gamma_K^{(i)}$ 
depends implicitly on the colour representation of parton $i$. To solve 
the equations we need to make this dependence explicit. Given that 
$\gamma_K^{(i)}$ admits Casimir scaling (\ref{Casimir_scaling_K}) 
at least to three loops, we write
\begin{equation}
\label{gamma_K_general}
 \gamma_K^{(i)} \left( \alpha_s \right) \equiv C_{i} \, 
 \, \widehat{\gamma}_K \left( \alpha_s \right) + 
 \widetilde{\gamma}_K^{(i)} \, ,
\end{equation}
where $\widetilde{\gamma}_K^{(i)} = {\cal O}(\alpha_s^4)$ accounts for
possible dependence on the representation of parton $i$ through higher-order
Casimir operators. It is presently an open question\footnote{An argument 
against Casimir scaling has been made~\cite{Juan,Adi}, based on the 
dependence of $\gamma_K^{(i)}$ on the representation in the strong coupling 
limit at large $N_c$. The argument is based on a class of antisymmetric representations with $k$ indices, where the ratio $k/N_c$ is kept fixed 
when $N_c\to \infty$; in this case the strong coupling limit of 
$\gamma_K^{(i)}$ does not admit Casimir scaling.} whether such terms 
appear.

Our constraints now take the form
\begin{eqnarray}
\label{oureq_reformulated1}
 \sum_{j \neq i} \frac{\partial}{\partial \ln(\rho_{i j})} \,
 \Gamma^{{\overline{\cal S}}} \left( 
 \rho_{i j}, \alpha_s \right) & = & \frac{1}{4} \, \bigg[C_i  \,
 \widehat{\gamma}_K \left( \alpha_s \right) \,+\, 
 \widetilde{\gamma}_K^{(i)} \left( \alpha_s \right)\bigg] \, ,
 \qquad \quad \forall i \, .
\end{eqnarray}
Using the linearity of these equations we can obviously write the general 
solution as a superposition of two functions
\begin{equation}
 \Gamma^{{\overline{\cal S}}}=\Gamma^{{\overline{\cal S}}}_{{\text{Q.C.}}}
 + \Gamma^{{\overline{\cal S}}}_{{\text{H.C.}}}
\end{equation}
which are, respectively, solutions of the equations
\begin{align}
\label{oureq_reformulated_QC}
\sum_{j \neq i} \frac{\partial}{\partial \ln(\rho_{i j})} \,
\Gamma^{{\overline{\cal S}}}_{{\text{Q.C.}}} \left( 
\rho_{i j}, \alpha_s \right) & =  \frac{1}{4} \, 
{\rm T}_i \cdot {\rm T}_i \,\widehat{\gamma}_K \left( \alpha_s \right) \, ,\qquad\qquad \forall i\,,
\\
\label{oureq_reformulated_HC}
\sum_{j \neq i} \frac{\partial}{\partial \ln(\rho_{i j})} \,
\Gamma^{{\overline{\cal S}}}_{{\text{H.C.}}} \left( 
\rho_{i j}, \alpha_s \right) & =  \frac{1}{4} \,  \widetilde{\gamma}_K^{(i)}\left( \alpha_s \right)\,, \qquad\qquad \forall i\,.
\end{align}
Here {\text{Q.C.}} and {\text{H.C.}} stand for Quadratic Casimir and 
Higher-order Casimir, respectively. Let us now focus on determining 
$\Gamma^{{\overline{\cal S}}}_{{\text{Q.C.}}}$, leaving aside 
$\Gamma^{{\overline{\cal S}}}_{{\text{H.C.}}} = {\cal O}(\alpha_s^4)$, 
which will be briefly discussed in sec.~\ref{sec:beyond}.

A solution for $\Gamma^{{\overline{\cal S}}}$, obeying 
eq.~(\ref{oureq_reformulated_QC}), is given by
\begin{equation}
\label{ansatz}
 \Gamma^{\overline{S}}
 \left(\rho_{i j}, \alpha_s \right) \, = \, - \frac18 \,
 \widehat{\gamma}_K\left(\alpha_s \right)  
 \sum_{(i,j)} \, \ln(\rho_{ij}) \, \mathrm{T}_i \cdot  \mathrm{T}_j 
 \, + \, \frac12 \, \widehat{\delta}_{{\overline{\cal S}}} ( \alpha_s )  
 \sum_{i = 1}^n  \mathrm{T}_i \cdot \mathrm{T}_i \, ,
\end{equation} 
where $\sum_{(i,j)}$ in the first term in (\ref{ansatz}) indiactes a sum over all pairs of hard partons, forming a colour dipole; each dipole is counted twice in the sum. Note this term carries the entire dependence on kinematics, correlating it with the colour structure. In contrast, the second term is independent of kinematics and is proportional to the unit matrix in colour space. Analysis of the $n = 2$ case (the Sudakov form factor) allows to identify this function (see eqs. (4.7) and (4.9) in~\cite{Gardi:2009qi}) as the one appearing at single-pole level in the partonic jet, the last term in eq.~(\ref{J_explicit}). 

It is easy to verify that (\ref{ansatz}) satisfies (\ref{oureq_reformulated_QC}): taking a derivative with respect 
to $\ln (\rho_{ij})$, for specific partons $i$ and $j$, isolates the color dipole
$\mathrm{T}_i \cdot  \mathrm{T}_j$; summing over $j$ for fixed $i$, and
enforcing colour conservation, given by eq.~(\ref{T_prop2}), one recovers
eq.~(\ref{oureq_reformulated_QC}). 

Integrating the renormalization group equation (\ref{rencalS}), with 
$\Gamma^{\overline{S}}$ given by eq.~(\ref{ansatz}), we obtain
an expression for the reduced soft function,
\begin{align}
\begin{split}
\label{barS_ansatz}
\overline{S}
  \left(\rho_{i j}, \alpha_s,\epsilon\right) = \, \exp\Bigg\{
  - \frac12 \int_0^{\mu^2}\frac{d\lambda^2}{\lambda^2}\Bigg[
  &\, \frac12 \, \, \widehat{\delta}_{{\overline{\cal S}}} ( 
  \alpha_s(\lambda^2,\epsilon) )  \, 
  \sum_{i = 1}^n  \mathrm{T}_i \cdot \mathrm{T}_i
  \\&  - \frac18 \,
  \widehat{\gamma}_K\left(\alpha_s(\lambda^2,\epsilon) \right)   
  \sum_{(i,j)} \,
  \ln \left( \rho_{i j}
  \right) 
\,  \mathrm{T}_i \cdot   \mathrm{T}_j\, 
\Bigg]\Bigg\}  \,.
\end{split}
\end{align}
Substituting eq.~(\ref{barS_ansatz}) into the factorization formula, 
eq.~(\ref{facamp}), together with the corresponding expression for the partonic 
jet, eq.~(\ref{J_explicit}), we obtain a complete description of
the singularity structure of the amplitude, eq.~(\ref{Z}).
Note that the $Z$ factor and the hard amplitude $\cal H$ in~(\ref{introducing_Z}) are 
separately independent of the auxiliary vectors $n_i$, as they must be. In 
contrast, the various elements in the factorization formula (\ref{facamp}) 
do depend on these vectors.
The cancellation of this dependence is non-trivial: it is guaranteed by the 
fact that $\overline{\cal S}$ admits the constraints of~(\ref{constraints}), 
and by the fact that the kinematic dependence of the singularities of the 
partonic jet function (\ref{J_explicit}) matches the one of the eikonal jet, 
eq.~(\ref{fincalJ}). It is essential that all the single pole terms 
that carry $n_i$ dependence in the various functions are governed by the 
cusp anomalous dimension $\gamma_K$ alone. Indeed, to obtain eq.~(\ref{Z}), 
we combine terms proportional to $\gamma_K$ in the soft and jet functions. 
In doing so we use colour conservation, $\sum_{j \neq i} \mathrm{T}_j = 
- \mathrm{T}_i$, as well as the relation between the kinematic variables 
of the various functions,
\begin{equation}
\label{kinematic_variables_combined}
\displaystyle{\underbrace{\ln \left(\frac{(2p_i\cdot n_i)^2}{n_i^2}
\right)}_{J_i}\,+\,
\underbrace{\ln \left(\frac{(2p_j\cdot n_j)^2}{n_j^2}\right)}_{J_j}}\,
+ \, \underbrace{\ln \left(\frac{\left(\beta_i\cdot\beta_j \, 
\,{\rm e}^{{\rm i} \pi\lambda_{ij}}\right)^2}{
\displaystyle{\frac{2(\beta_i\cdot n_i)^2}{n_i^2}\frac{2(\beta_j
\cdot n_j)^2}{n_j^2}}}\right)}_{\overline{\cal S}}
= 2 \ln (2 p_i\cdot p_j\,{\rm e}^{{\rm i} \pi\lambda_{ij}}) .
\end{equation}
Note also that the poles associated with $\widehat{\delta}_{{\overline{\cal S}}}
(\alpha_s(\lambda^2,\epsilon))$  cancel out between the soft and jet functions, 
given by eqs.~(\ref{barS_ansatz}) and (\ref{J_explicit}), respectively. 

It is interesting to compare at this point our approach
to that of Becher and Neubert in Ref.~\cite{Becher:2009qa}. The final 
expression at the amplitude level, eq.~(\ref{Z}), is the same. The set of 
constraints, eq.~(48) in~\cite{Becher:2009qa}, is also equivalent. The 
underlying factorization scheme, and consequently the arguments leading 
to these constraints, are however somewhat different. In particular, 
ref.~\cite{Becher:2009qa} does not define jet functions using auxiliary 
Wilson lines ($n_i$ in our formulation); instead, it keeps track of the jets 
through their mass, taking $p_i$ slightly off the light cone, $p_i^2\neq0$. 
In their formulation, the equivalent of eq.~(\ref{kinematic_variables_combined}) 
takes the form (eq.~(43) in \cite{Becher:2009qa})
\begin{equation}
\label{kinematic_variables_BN}
 \displaystyle{\underbrace{\ln \left(\frac{-p_i^2}{\mu^2}
 \right)}_{J_i}\,+\,
 \underbrace{\ln \left(\frac{-p_j^2}{\mu^2}
 \right)}_{J_j}}\, 
 \, + \, \underbrace{\ln \left(\frac{
 2 \, p_i \cdot p_j \, {\rm e}^{{\rm i} \pi\lambda_{ij}}
 \mu^2}{(- p_i^2)(- p_j^2)} \right)}_{S} \, = \ 
 \ln \left(\frac{2 p_i\cdot p_j\,{\rm e}^{{\rm i} 
 \pi\lambda_{ij}}}{\mu^2}\right) \, ,
\end{equation} 
which is again realised owing to the fact that in each function the corresponding 
logarithm is governed by the cusp anomalous dimension. Recall that in our
derivation the argument of the reduced soft function is dictated by rescaling
invariance; in contrast, in \cite{Becher:2009qa} the argument of the soft 
function is essentially dictated by power counting, and it is not invariant with
respect to rescaling.

\section{Possible contributions beyond the sum-over-dipoles formula
\label{sec:beyond}}

As already mentioned, the ansatz of (\ref{ansatz}) does not in general 
provide a unique solution of the available constraints. Thus eq.~(\ref{Z}), 
while consistent with all existing calculations, may still be missing some 
large-angle soft singularities beyond a certain loop order. Such further
singularities are however strongly constrained both in their functional form
and in their color structure. It is worthwhile emphasizing that calculations in 
the large--$N_c$ limit cannot resolve the question at hand, since planar 
eikonal diagrams necessarily factorize into a product of colour dipoles, those 
made of adjacent Wilson lines, and are therefore automatically consistent 
with the sum-over-dipoles formula. The analysis must therefore be done 
at finite~$N_c$. Ref.~\cite{Gardi:2009qi} has identified two classes of 
corrections that may appear, and although some progress was made, neither 
of the two can be excluded to all orders at present. 

The first class of corrections corresponds to potential higher-order Casimir
contributions. In case higher-order Casimir operators do show up in the cusp
anomalous dimension at some loop order, {\it i.e.} $\widetilde{\gamma}_K$ in 
(\ref{gamma_K_general}) does not vanish, the anomalous dimension of the
reduced soft function of any amplitude will receive additional corrections.  
These corrections are subject to the very stringent constraints of 
(\ref{oureq_reformulated_HC}). For amplitudes with two or three legs 
these corrections still have a dipole structure (see {\it e.g.} Appendix A 
in~\cite{Gardi:2009qi}), however, for amplitudes with four legs or more, 
non-trivial structures that couple more than two hard patrons may arise.
 
The second class of corrections, which may be present even if ${\gamma}_K$
admits Casimir scaling, is given by solutions of the homogeneous equation 
associated with eq.~(\ref{oureq_reformulated_QC}). Indeed, adding to our 
ansatz any function $\Delta^{\overline{S}} (\rho_{ij})$ satisfying
\begin{equation}
\label{Delta_oureq_reformulated}
\sum_{j \neq i} \frac{\partial}{\partial \ln(\rho_{i j})} 
\Delta^{{\overline{\cal S}}} \left( 
\rho_{i j}, \alpha_s \right) =  0 \, \qquad \forall i \, ,
\end{equation}
one obtains a new solution of eq.~(\ref{oureq_reformulated_QC}).
Eq.~(\ref{Delta_oureq_reformulated}) is solved by any function of the 
conformal invariant cross ratios defined in (\ref{rhoijkl}). Any such solution
has the property of being invariant with respect to velocity rescalings 
without involving the jets. Such functions can of course be written directly 
in terms of the original kinematic variables $p_i\cdot p_j$ and are therefore 
not constrained by soft--collinear factorization.  

Interesting examples for $\Delta^{{\overline{\cal S}}}$ in the four parton case were proposed in~\cite{Gardi:2009qi}:
\begin{align}
\label{hat_H}
&\sum_{j,k,l} \sum_{a,b,c}
{\rm i} \, f_{abc} \,{\rm T}_j^{a} {\rm T}_k^{b} {\rm T}_l^{c} \,
\ln \left(\rho_{ijkl}\right) \, \ln \left(\rho_{iklj}\right) \,
\ln \left(\rho_{iljk}\right) \,, \\
\label{hat_H_}
&\sum_{j,k,l} \sum_{a,b,c}
d_{abc} \, {\rm T}_j^{a} {\rm T}_k^{b} {\rm T}_l^{c} \,
\ln^2 \left(\rho_{ijkl}\right) \, \ln^2 \left(\rho_{iklj}\right) \,
\ln^2 \left(\rho_{iljk}\right) \, ,
\end{align}
where the sum over partons is understood to exclude identical indices, and where colour conservation, ${\rm T}_i^{d}=-{\rm T}_j^{d} -{\rm T}_k^{d} -{\rm T}_l^{d}$,  has been taken into account. 
Note that these functions are, by contruction, symmetric 
under the exchange of Wilson lines (Bose symmetry): this correlates colour 
and kinematic degrees of freedom. These functions, moreover, do not contribute 
in the limit where any two hard partons become collinear, and therefore they 
cannot be excluded using the properties of the splitting amplitude discussed in~\cite{Becher:2009qa}.

Functions of conformal-invariant cross ratios such as (\ref{hat_H}) correlate colour and kinematic degrees of freedom of four partons. They cannot arise at two loops because two-loop webs can connect at most three partons. This explains, a posteriori, the findings of ref.~\cite{MertAybat:2006mz}, which explicitly showed that there are no new correlations generated at the two loop order beyond those 
of pairwise interactions\footnote{Note that new structure does appear in the case 
of scattering involving heavy quarks, as shown in 
refs.~\cite{Mitov:2009sv,Kidonakis:2009ev,Becher:2009kw,Gluza:2009yy,Ferroglia:2009ep,Czakon:2009zw,Beneke:2009rj}.}. 
Non-trivial corrections to the the sum-over-dipoles formula can therefore first 
arise at three loops.

Unfortunately, at three loops no complete calculation is available yet. However, several important steps have been taken. First, as already emphasized in 
\cite{Gardi:2009qi}, three-loop corrections to this formula must satisfy 
eq.~(\ref{Delta_oureq_reformulated}) --- they must be functions of conformal 
invariant cross ratios, and they must vanish identically in amplitudes of less than 
four legs. Beyond that, it was explicitly shown that the class of three-loop 
diagrams containing matter loops is consistent with the sum-over-dipoles 
formula~\cite{Dixon:2009gx}.  

A further step was taken in Ref.~\cite{Becher:2009qa}, where it was shown that if dependence 
on the kinematic variables is assumed to be single--logarithmic, there is 
no possible structure that could appear at three-loops beyond the 
sum-over-dipoles formula. This argument is based on eliminating all possible 
structures using the factorization constraints discussed above, together with 
Bose symmetry, and an additional constraint on the singularity structure in 
the limit where two hard partons become collinear based on the properties of the splitting amplitude. It should be emphasized that the assumption of single--logarithmic kinematic dependence is crucial for this argument, and this assumption may well be violated. It thus remains an open question whether new structures appear in the soft anomalous dimension at three loops.

\section{Conclusions}

We have reviewed recent exciting progress in determining the infrared 
singularities of on-shell scattering amplitudes in massless non-Abelian 
gauge theories. 
It is now firmly established~\cite{Gardi:2009qi,Becher:2009qa} that the cusp
anomalous dimension has a central role in governing soft singularities of 
multi-leg amplitudes with an arbitrary number of legs and for a general~$N_c$. 
This role is summarized by a set of differential equations (\ref{constraints}) 
constraining the kinematic dependence of the soft anomalous dimension matrix 
of any amplitude, to any loop order, in an arbitrary colour basis. These constraints are a direct consequence of factorization and of the special properties of soft 
gluon interactions with massless hard partons.

The simplest solution to this set of constraints yields a closed form expression 
for the singularities of any massless scattering amplitude, eq.~(\ref{Z}).  
According to this formula the correlations induced by soft gluon interactions 
between colour and kinematic degrees of freedom take the form of a sum 
over colour dipoles. No new correlations are generated by multi-loop webs 
(fig.~\ref{fig:webs}): the colour matrix structure remains the same as at 
one loop, and the cusp anomalous dimension alone governs all non-collinear
singularities.

We have further shown that possible corrections to this simple sum-over-dipoles 
formula belong to one of two categories: ones that are generated by potential
higher-order Casimir contributions to the cusp anomalous dimension, which must
then satisfy eq.~(\ref{oureq_reformulated_HC}), and ones that can be written in 
terms of conformal invariant cross ratios (\ref{rhoijkl}), solving the homogeneous 
equations (\ref{Delta_oureq_reformulated}). The former may contribute to any 
amplitude starting from four loops, while the latter can only appear in amplitudes 
with four or more hard partons, starting at three loops. So far all explicit 
calculations are consistent with the sum-over-dipoles formula, but it remains 
an open question whether such corrections do show up at some loop order.

\acknowledgments

We would like to thank Lance Dixon, Juan Maldacena, Gregory Korchemsky 
and George Sterman for useful discussions. Work supported in part by the 
European Community's Marie-Curie Research Training Network `Tools and 
Precision Calculations for Physics Discoveries at Colliders'  (`HEPTOOLS'), 
contract MRTN-CT-2006-035505.


\begin{thebibliography}{0}

\bibitem{Bern:2005iz}
  Z.~Bern, L.~J.~Dixon and V.~A.~Smirnov,
  Phys.\ Rev.\  D {\bf 72} (2005) 085001, hep-th/0505205.

\bibitem{Bern:2008ap}
  Z.~Bern, L.~J.~Dixon, D.~A.~Kosower, R.~Roiban, M.~Spradlin, C.~Vergu 
  and A.~Volovich,
  Phys.\ Rev.\  D {\bf 78} (2008) 045007, arXiv:0803.1465 [hep-th].

\bibitem{Korchetal}
  J.~M.~Drummond, G.~P.~Korchemsky and E.~Sokatchev,
  \NP{B 795} (2008) 385,  arXiv:0707.0243 [hep-th];
  A.~Brandhuber, P.~Heslop and G.~Travaglini,
  \NP{B 794} (2008) 231,  arXiv:0707.1153 [hep-th];
  J.~M.~Drummond, J.~Henn, G.~P.~Korchemsky and E.~Sokatchev,
  \NP{B 795} (2008) 52, arXiv:0709.2368 [hep-th];
  arXiv:0712.1223 [hep-th];
  Phys.\ Lett.\  B {\bf 662} (2008) 456, arXiv:0712.4138 [hep-th];
  Nucl.\ Phys.\  B {\bf 815} (2009) 142, arXiv:0803.1466 [hep-th]. 

\bibitem{Alday:2007hr}
  L.~F.~Alday and J.~M.~Maldacena,
  JHEP {\bf 0706} (2007) 064, arXiv:0705.0303 [hep-th].

\bibitem{Alday:2008yw}
  L.~F.~Alday and R.~Roiban,
  Phys.\ Rept.\  {\bf 468} (2008) 153, arXiv:0807.1889 [hep-th].

\bibitem{Polyakov:1980ca}
  A.~M.~Polyakov,
  Nucl.\ Phys.\  B {\bf 164} (1980) 171.

\bibitem{Korchemsky:1985xj}
  G.~P.~Korchemsky and A.~V.~Radyushkin,
  \PL{B 171} (1986) 459.

\bibitem{Ivanov:1985np}
  S.~V.~Ivanov, G.~P.~Korchemsky and A.~V.~Radyushkin,
  Yad.\ Fiz.\ {\bf 44} (1986) 230,
  [Sov.\ J.\ Nucl.\ Phys.\ {\bf 44} (1986) 145].

\bibitem{Korchemsky:1987wg}
  G.~P.~Korchemsky and A.~V.~Radyushkin, \NP{B 283} (1987) 342.

\bibitem{Korchemsky:1988hd}
  G.~P.~Korchemsky, \PL{B 220} (1989) 629.

\bibitem{Korchemsky:1988si}
  G.~P.~Korchemsky,
  Mod.\ Phys.\ Lett.\  A {\bf 4} (1989) 1257.

\bibitem{Moch:2004pa}
  S.~Moch, J.~A.~M.~Vermaseren and A.~Vogt,
  Nucl.\ Phys.\  B {\bf 688} (2004) 101, hep-ph/0403192.

\bibitem{Beisert:2005fw}
  N.~Beisert and M.~Staudacher,
  Nucl.\ Phys.\  B {\bf 727} (2005) 1, hep-th/0504190.

\bibitem{Beisert:2006ez}
  N.~Beisert, B.~Eden and M.~Staudacher,
  J.\ Stat.\ Mech.\  {\bf 0701} (2007) P021, hep-th/0610251.

\bibitem{Basso:2007wd}
  B.~Basso, G.~P.~Korchemsky and J.~Kotanski,
  Phys.\ Rev.\ Lett.\  {\bf 100} (2008) 091601, arXiv:0708.3933 [hep-th].

\bibitem{Bloch:1937pw}
  F.~Bloch and A.~Nordsieck,
  Phys.\ Rev.\  {\bf 52} (1937) 54.

\bibitem{Kinoshita:1962ur}
  T.~Kinoshita,
  J.\ Math.\ Phys.\  {\bf 3} (1962) 650.

\bibitem{Lee:1964is}
  T.~D.~Lee and M.~Nauenberg,
  Phys.\ Rev.\  {\bf 133} (1964) B1549.

\bibitem{Catani:1996jh}
  S.~Catani and M.~H.~Seymour,
  Phys.\ Lett.\  B {\bf 378} (1996) 287, hep-ph/9602277.

\bibitem{Gardi:2001ny}
  E.~Gardi and J.~Rathsman,
  Nucl.\ Phys.\  B {\bf 609} (2001) 123, hep-ph/0103217; 
  Nucl.\ Phys.\  B {\bf 638} (2002) 243, hep-ph/0201019.

\bibitem{Cacciari:2002xb}
  M.~Cacciari and E.~Gardi,
  Nucl.\ Phys.\  B {\bf 664} (2003) 299, hep-ph/0301047.

\bibitem{Bozzi:2007pn}
  G.~Bozzi, S.~Catani, D.~de Florian and M.~Grazzini,
  Nucl.\ Phys.\  B {\bf 791} (2008) 1, arXiv:0705.3887 [hep-ph].

\bibitem{Ahrens:2008nc}
  V.~Ahrens, T.~Becher, M.~Neubert and L.~L.~Yang,
  arXiv:0809.4283 [hep-ph].

\bibitem{Andersen:2005mj}
  J.~R.~Andersen and E.~Gardi,
  JHEP {\bf 0601} (2006) 097, hep-ph/0509360.

\bibitem{Andersen:2006hr}
  J.~R.~Andersen and E.~Gardi,
  JHEP {\bf 0701} (2007) 029, hep-ph/0609250;
  JHEP {\bf 0506} (2005) 030, hep-ph/0502159.

\bibitem{Gardi:2006jc}
  E.~Gardi,
    ``Inclusive distributions near kinematic thresholds,''
  {\it In the Proceedings of FRIF workshop on first principles non-perturbative 
  QCD   of hadron jets, LPTHE, Paris, France, 12-14 Jan 2006, pp E003}
  hep-ph/0606080.

\bibitem{Sterman:1986aj}
  G.~Sterman,
  Nucl.\ Phys.\  B {\bf 281} (1987) 310.

\bibitem{Catani:1989ne}
  S.~Catani and L.~Trentadue,
  Nucl.\ Phys.\  B {\bf 327} (1989) 323.

\bibitem{Contopanagos:1996nh}
  H.~Contopanagos, E.~Laenen and G.~Sterman,
  Nucl.\ Phys.\  B {\bf 484}, 303 (1997), hep-ph/9604313.

\bibitem{Laenen:2000ht}
  E.~Laenen, G.~Sterman and W.~Vogelsang, hep-ph/0010184.

\bibitem{Gardi:2007ma}
  E.~Gardi and G.~Grunberg,
  Nucl.\ Phys.\  B {\bf 794}, 61 (2008), arXiv:0709.2877 [hep-ph].

\bibitem{Gardi:2002xm}
  E.~Gardi and R.~G.~Roberts,
  Nucl.\ Phys.\  B {\bf 653} (2003) 227, hep-ph/0210429.

\bibitem{Catani:1992ua}
  S.~Catani, L.~Trentadue, G.~Turnock and B.~R.~Webber,
  Nucl.\ Phys.\  B {\bf 407} (1993) 3.

\bibitem{Kidonakis:1999ze}
  N.~Kidonakis,
  Int.\ J.\ Mod.\ Phys.\  A {\bf 15} (2000) 1245, hep-ph/9902484.

\bibitem{Becher:2006mr}
  T.~Becher, M.~Neubert and B.~D.~Pecjak,
  JHEP {\bf 0701} (2007) 076, hep-ph/0607228.

\bibitem{Botts:1989kf}
  J.~Botts and G.~Sterman,
  Nucl.\ Phys.\  B {\bf 325} (1989) 62.

\bibitem{Kidonakis:1997gm}
  N.~Kidonakis and G.~Sterman,
  Nucl.\ Phys.\  B {\bf 505} (1997) 321, hep-ph/9705234.

\bibitem{MertAybat:2006mz}
  S.~Mert Aybat, L.~J.~Dixon and G.~Sterman,
  Phys.\ Rev.\  D {\bf 74} (2006) 074004, hep-ph/0607309.

\bibitem{Gardi:2009qi}
  E.~Gardi and L.~Magnea,
  JHEP {\bf 0903} (2009) 079, arXiv:0901.1091 [hep-ph].

\bibitem{Becher:2009cu}
  T.~Becher and M.~Neubert, arXiv:0901.0722 [hep-ph].

\bibitem{Becher:2009qa}
  T.~Becher and M.~Neubert, arXiv:0903.1126 [hep-ph].

\bibitem{Catani:1998bh}
  S.~Catani,
  Phys.\ Lett.\  B {\bf 427} (1998) 161, hep-ph/9802439.

\bibitem{Magnea:1990zb}
  L.~Magnea and G.~Sterman,
  Phys.\ Rev.\  D {\bf 42} (1990) 4222.

\bibitem{Magnea:2000ss}
  L.~Magnea,
  Nucl.\ Phys.\  B {\bf 593} (2001) 269, hep-ph/0006255.

\bibitem{Vogt:2004mw}
  A.~Vogt, S.~Moch and J.~A.~M.~Vermaseren,
  Nucl.\ Phys.\  B {\bf 691} (2004) 129, hep-ph/0404111.

\bibitem{Moch:2005id}
  S.~Moch, J.~A.~M.~Vermaseren and A.~Vogt,
  JHEP {\bf 0508} (2005) 049, hep-ph/0507039.

\bibitem{Moch:2005tm}
  S.~Moch, J.~A.~M.~Vermaseren and A.~Vogt,
  Phys.\ Lett.\  B {\bf 625} (2005) 245, hep-ph/0508055.

\bibitem{Sterman:2002qn}
  G.~Sterman and M.~E.~Tejeda--Yeomans, 
  \PL{B 552} (2003) 48, hep-ph/0210130. 

\bibitem{Dixon:2008gr}
  L.~J.~Dixon, L.~Magnea and G.~Sterman,
  JHEP {\bf 0808} (2008) 022, arXiv:0805.3515 [hep-ph].

\bibitem{Sen:1982bt}
  A.~Sen, \PR{D 28} (1983) 860.

\bibitem{Kidonakis:1998nf}
  N.~Kidonakis, G.~Oderda and G.~Sterman,
  Nucl.\ Phys.\  B {\bf 531} (1998) 365, hep-ph/9803241.

\bibitem{Dotsenko:1979wb}
  V.~S.~Dotsenko and S.~N.~Vergeles,
  Nucl.\ Phys.\  B {\bf 169} (1980) 527.

\bibitem{Brandt:1981kf}
  R.~A.~Brandt, F.~Neri and M.~a.~Sato,
  Phys.\ Rev.\  D {\bf 24} (1981) 879.

\bibitem{Korchemskaya:1994qp}
  I.~A.~Korchemskaya and G.~P.~Korchemsky,
  Nucl.\ Phys.\  B {\bf 437} (1995) 127, hep-ph/9409446.

\bibitem{Juan}
Juan Maldacena, private communication, January 2009.  

\bibitem{Adi}
  A.~Armoni,
  JHEP {\bf 0611} (2006) 009, hep-th/0608026.

\bibitem{Mitov:2009sv}
  A.~Mitov, G.~Sterman and I.~Sung, arXiv:0903.3241 [hep-ph].

\bibitem{Kidonakis:2009ev}
  N.~Kidonakis, arXiv:0903.2561 [hep-ph].

\bibitem{Becher:2009kw}
  T.~Becher and M.~Neubert, arXiv:0904.1021 [hep-ph].

\bibitem{Gluza:2009yy}
  J.~Gluza, A.~Mitov, S.~Moch and T.~Riemann, arXiv:0905.1137 [hep-ph].

\bibitem{Beneke:2009rj}
  M.~Beneke, P.~Falgari and C.~Schwinn,
  arXiv:0907.1443 [hep-ph].

\bibitem{Czakon:2009zw}
  M.~Czakon, A.~Mitov and G.~Sterman,
  arXiv:0907.1790 [hep-ph].

\bibitem{Ferroglia:2009ep}
  A.~Ferroglia, M.~Neubert, B.~D.~Pecjak and L.~L.~Yang,
  arXiv:0907.4791 [hep-ph].


\bibitem{Dixon:2009gx}
  L.~J.~Dixon,
  {\em Phys. Rev.} {\bf D}79, 091501,  arXiv:0901.3414 [hep-ph].



\end{thebibliography}
\end{document}